\documentclass[12pt]{iopart}
\input epsf
\begin{document}
\jl{2}

\title{The Moment of Inertia and the Scissors Mode of a Bose-condensed Gas}
\author{O.M. Marag\`o, G. Hechenblaikner, E. Hodby, S.A. Hopkins and
C.J. Foot}
\address{Clarendon Laboratory, Department of Physics, University of Oxford,\\
Parks Road, Oxford, OX1 3PU, United Kingdom.}
\date{\today}


\begin{abstract}
We relate the frequency of the scissors mode to the moment of
inertia of a trapped Bose gas at finite temperature in a
semi-classical approximation. We apply these theoretical results
to the data obtained in our previous study of the properties of
the scissors mode of a trapped Bose-Einstein condensate of
$^{87}$Rb atoms as a function of the temperature. The frequency
shifts that we measured show quenching of the moment of inertia of
the Bose gas at temperatures below the transition temperature -
the system has a lower moment of inertia that of a rigid body with
the same mass distribution, because of superfluidity.
\end{abstract}

\pacs{03.75.Fi, 05.30.Jp, 32.80.Pj, 67.90.+z}

\section{INTRODUCTION}
The onset of Bose-Einstein condensation~\cite{Anderson} (BEC) in
ultra-cold atomic gases is clearly observed through the change in
the density profile of the confined system. On the contrary
superfluid behavior in these systems has proved to be more
difficult to observe. Four years after the first observation of
BEC in atomic gases, evidence for superfluidity was `still quite
circumstantial'~\cite{Leggett99} but in recent years superfluid
phenomena in dilute gases have been the subject of much
theoretical and experimental work~\cite{Timmermans2001}.
Superfluidity in trapped condensates was clearly demonstrated by
experiments that showed evidence for a critical
velocity~\cite{Raman99}, the creation of
vortices~\cite{Matthews1999a,Madison2000a} and the observations of
the scissors mode~\cite{David99,Marago2000}. Many striking
features of superfluids are associated with their response to
rotation. In this paper we concentrate only on one aspect, namely
the quenching of the moment of inertia, so that the superfluid
always has less moment of inertia than a rigid body with same
density distribution. We first review our recent results on the
scissors mode of excitation, which is our tool to measure the
moment of inertia of the Bose gas. We then use the theory recently
developed by Zambelli and Stringari~\cite{Zambelli2001} to deduce
the moment of inertia of the gas at finite temperature.

\section{THE SCISSORS MODE}
\label{scistheory}

The scissors mode of excitation was first studied in atomic nuclei
and predicted by a geometrical model~\cite{nicola}. Its
experimental discovery~\cite{Bohle} has been one of the most
exciting findings in nuclear physics during the last two
decades~\cite{review}. According to  the geometrical picture, such
a mode arises from a counter-rotational oscillation of the
deformed proton and neutron fluids, where the axes of the deformed
clouds move like the blades of a pair of scissors. Extensive
studies of this mode have investigated important features such as
the dependence of its strength on the nuclear deformation and its
relation with the superfluid moment of inertia of the
nucleus~\cite{review}.

In a theoretical paper~\cite{David99} D. Gu\'ery-Odelin and S.
Stringari extended the study of the scissors mode to the case of
trapped gases. They investigated the oscillatory excitation of a
dilute atomic gas following a sudden rotation of the anisotropic
confining potential and they showed how the scissors mode is a
manifestation of the superfluidity of a trapped BEC. In this
section we apply their approach to the specific geometry of our
TOP trap and discuss the expected results for the scissors mode of
both the BEC and the classical gas. We then show experimental
results for the scissors mode oscillations and finally put the
scissors mode into the context of the low-energy spectrum of BEC
excitations.

A Bose condensate of density $n(\vec{r},t)$ at $T=0$ is well
described by the hydrodynamic equations of superfluids:
\begin{eqnarray}
\frac{\partial n}{\partial t} + \nabla \cdot (n \vec{v}) =0&\ \ \
\ & {\rm Continuity\ Equation} \label{continuity}\\
m\frac{\partial \vec{v}}{\partial t} + \nabla \left(V_{ext} +gn +
\frac{m v^2}{2}\right)=0 &\ \ \ \ & {\rm Force\ Equation}
\label{force}
\end{eqnarray}
where $V_{ext}$ is the external potential, $g$ characterizes the
particle interaction strength and $m$ is the mass of the
particles. The velocity flow $\vec{v}(\vec{r},t)$ is related to
the phase $S(\vec{r},t)$ of the complex order parameter
$\Phi=\sqrt{n}\, e^{iS}$ through:
\begin{equation}
\vec{v}(\vec{r},t)=\frac{\hbar}{m}\nabla
S(\vec{r},t).\label{theory-vel}
\end{equation}
This relation immediately leads to $\nabla\times \vec{v}=0$, i.e.
the velocity flow is irrotational. Since $\Phi$ must be
single-valued the phase change over a closed path must be modulo
$2\pi$. In the case of a BEC with a large number of atoms, we can
use the Thomas-Fermi ground state density distribution $n= [ \mu
-V_{ext}(\vec{r})]/g$ as a stationary
solution~\cite{Dalfovo1999a}, where $\mu$ is the chemical
potential.

The gas is initially in thermal equilibrium in an anisotropic
harmonic potential $V_{ext}(r)$ with three frequencies $\omega_x
\simeq \omega_y < \omega_z$
\begin{equation}
V_{ext}(r)=\frac{m}{2} (\omega_x^2 x^2 + \omega_y^2 y^2 +
\omega_z^2 z^2).\label{potential}
\end{equation}
It is convenient to define the two quantities
$\omega_{sc}=\sqrt{\omega_x^2 + \omega_z^2}$ and
$\epsilon=(\omega_z^2 - \omega_x^2)/(\omega_z^2 + \omega_x^2)$
where $\epsilon$ gives a measure of the deformation of the
confining potential. For the TOP trap case we have
$\omega_z=\sqrt{8}\,\omega_x$, so that $\omega_{sc}= 3\,\omega_x$
and $\epsilon=7/9$. To start the scissors mode at a certain time
one rotates the trapping potential through a small angle
$\theta_0$ about the y axis. The condensate is no longer in
equilibrium and the atomic cloud will start oscillating about the
y axis. After the sudden rotation of the trapping potential the TF
density can be rewritten with respect to axes of the rotated
potential. Making the small angle approximations $x\rightarrow
x-\theta z$ and $z\rightarrow z+\theta x$, leads to a
time-dependent density:
\begin{equation}
n(\vec{r},t)=\frac{\mu}{g}- \frac{m}{2g}[\omega_x^2 x^2 +
\omega_y^2 y^2 + \omega_z^2 z^2+ 2 \omega_{sc}^2\epsilon \theta(t)
xz]
\end{equation}
In the subsequent motion, the cloud is not deformed provided that
the change in the potential is too small to excite shape
oscillations. In other words the scissors motion is completely
decoupled from `compressional' modes and the velocity flow
satisfies the condition
\begin{equation}
\nabla \cdot \vec{v}=0.
\end{equation}
The velocity flow for a condensate is also irrotational. The
constraints on the velocity flow lead to
\begin{equation}
\vec{v}=\beta(t)\nabla (xz).
\end{equation}
Substituting the expressions for $n$ and $\vec{v}$ into
Eqs.~\ref{continuity} and \ref{force} we then get two equations
for the angle $\theta(t)$ and the phase parameter $\beta(t)$:
\begin{equation}
 \left \{
\begin{array}{lcl}
\dot{\theta}(t)& = & - \beta(t)/\epsilon\\ \dot{\beta}(t) & = &
\omega_{sc}^2\, \epsilon\, \theta(t)
\end{array}
\right. \label{scis-eqs}
\end{equation}
Oscillations cannot occur about an axis with cylindrical symmetry
i.e. $\epsilon=0$, since in this case the angle $\theta$ loses its
geometrical meaning and the phase of the condensate is constant.
When $\epsilon\neq 0$, Eqs.~\ref{scis-eqs} lead to harmonic
angular motion of the cloud $\ddot{\theta}=-\omega_{sc}^2\theta$.
The initial conditions $\theta (0)=\theta_0$ and $\vec{v}(0)=0$
yields the solutions:
\begin{equation}
 \left \{
\begin{array}{lcl}
\theta(t)& = & \theta_0 \cos (\omega_{sc}\, t)\\ \beta(t) & = &
\epsilon\, \theta_0\, \omega_{sc}\, \sin (\omega_{sc}\, t).
\end{array}
\right.
\end{equation}
The scissors mode is an oscillation of the condensed cloud at
$\omega_{sc}$ which is undamped at $T=0$. This result is a direct
consequence of superfluidity and in particular of the relation
between the velocity flow and the phase of the condensate. In this
sense the constraint on the overall phase of the order parameter
is transposed to the dynamical behavior of the cloud during a
scissors oscillation.

The thermal gas behaves in a completely different way after a
scissors excitation. In the so-called collisionless regime (when
the collision time is larger than the trap oscillation period),
the evolution of the cloud angle is the superposition of two
undamped oscillations at frequencies given by
$\omega_{\pm}=|\omega_z\pm \omega_x|$~\cite{David99}. The sudden
rotation of the potential excites both low and high frequency
oscillations with the same amplitude. These two modes correspond
to two types of flow pattern: the high frequency one is
irrotational flow whereas the low frequency oscillation is of
rotational nature, related to the rigid body value for the moment
of inertia of a classical gas. We can summarize these results by
writing the velocity flow of the classical gas in terms of a
rotational frequency $\Omega_{rot}$ and of a scalar
pseudo-potential $\chi(\vec{r})$ responsible for the irrotational
flow $\vec{v}=\vec{\Omega}_{rot} \times \vec{r} + \nabla
\chi(\vec{r})$ \cite{Lamb}.
\begin{figure}
\begin{center}\mbox{ \epsfxsize 5.5in\epsfbox{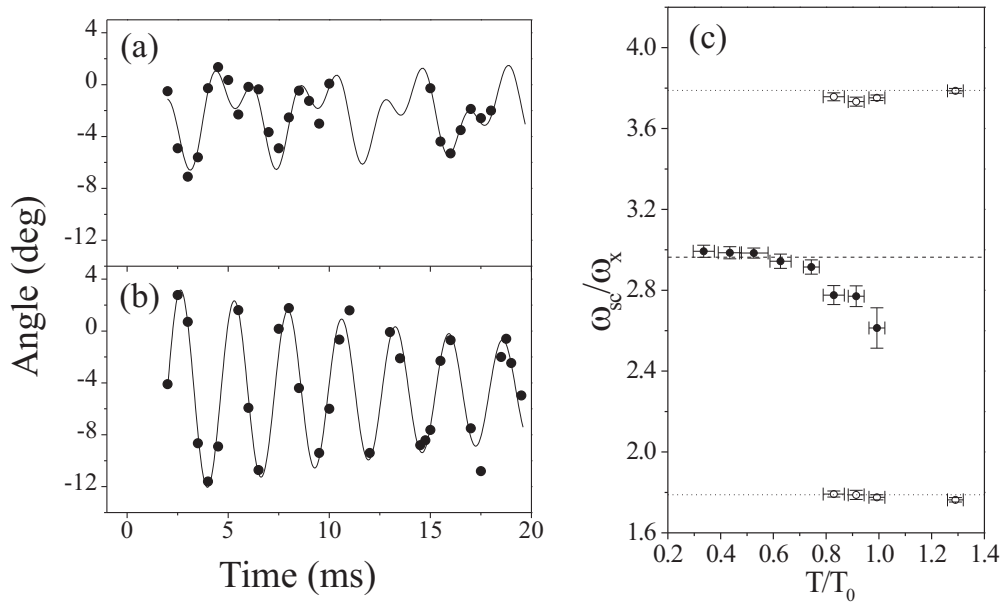}}\end{center}
\caption{Scissors mode oscillations and frequency shifts (as
reported in \cite{Marago2001}). (a) Scissors oscillation of a
thermal cloud at $T/T_0=1.29$. The scissors mode is characterized
by two frequencies of oscillation with equal amplitude. (b) A
condensate at $T/T_0=0.53$ exhibits an almost undamped oscillation
at a frequency that agrees well with the hydrodynamic value.  (c)
Temperature dependent frequency shift of the scissors mode of the
condensate (solid circles) and hydrodynamic prediction (dashed
line). The low temperature frequencies are systematically higher
than the hydrodynamic value by $\sim 1\%$ because of the finite
number of atoms. The scissors mode frequencies of the thermal
component are represented by open circles and the collisionless
predictions by the dotted lines. The frequency spectrum has been
normalized to $\omega_x$.} \label{freq}
\end{figure}

In our experiments~\cite{Marago2000,Marago2001} we use
radio-frequency induced evaporation to cool the atoms in a TOP
trap to the desired temperature $T$ (we obtain `pure' condensates
with $2\times 10^4$ atoms). The frequencies of this trap are
$\omega_x=\omega_y=2\pi \times 126\ {\rm Hz},\
\omega_z=\sqrt{8}\omega_x$. We then apply an oscillatory bias
field along the z axis, which adiabatically tilts the confining
potential by a small angle $\phi$ in $0.2$~s. This angle was kept
fixed at a value of $\phi=3.4^{\circ}$ for all the measurements.
This procedure also reduces $\omega_z$ by $\sim 2\%$. After the
adiabatic tilting, we suddenly flip the trap angle to $-\phi$.
This excites the scissors mode in the xz plane with an amplitude
$\theta_0=2\phi$, about the new equilibrium position. To observe
the scissors mode of the thermal cloud after the sudden rotation
of the potential, pictures of the atom cloud in the trap were
taken after a variable delay. The angle of the cloud was extracted
from a 2-D Gaussian fit of the absorption profiles. To observe the
scissors mode in a Bose-Einstein condensed gas we allow the BEC to
evolve in the trap for a variable time after exciting the scissors
mode and then release the condensate from the trap so that it
expands for $14$~ms before the time-of-flight image. The repulsive
mean-field interactions cause the cloud to expand rapidly when the
confining potential is switched off, so that its spread is much
greater than the initial size. The aspect ratio of the cloud
changes during the expansion (the long axis is always at
$90^{\circ}$ to the original long axis in the trap) but this does
not affect the results since we only extract the frequency at
which the cloud angle oscillates.

Figure~\ref{freq} shows in (a) the scissors mode oscillations for
the thermal cloud ($T/T_0=1.29$, $T_0$ being the transition
temperature) and in (b) for the condensate ($T/T_0=0.53$), in (c)
we plot the scissors mode frequency shift as a function of
temperature \cite{Marago2001}. The measured frequencies of the
thermal cloud do not change with temperature and agree very well
with the collisionless frequencies, both above and below the
critical temperature. Both frequency components have the same
amplitude implying that energy is shared equally between
rotational and irrotational velocity flow. The scissors mode
oscillation of the BEC component occurs at a {\it single
frequency} $\omega_{sc}$. However, the frequency of oscillation
and its damping change dramatically with
temperature~\cite{Marago2001}. Figure~\ref{freq}~(c) shows the
measured frequencies for the scissors mode of thermal cloud (open
circles) and condensate (solid circles) for different
temperatures. For the lower temperature BEC points we measure a
frequency that is systematically $\sim 1\%$ higher than the
hydrodynamic prediction. This agrees with the scissors mode
frequency we calculated for a finite number condensate ($2\times
10^4$) using the method by Pires and de Passos~\cite{Pires2000}.
Stimulated by our work, Jackson and
Zaremba~\cite{JacksonZaremba2001} have recently developed a
theoretical description of the scissors mode at finite $T$. Their
theoretical results are in good agreement with our experimental
results for a wide range of temperatures. Close to the critical
temperature, $T/T_0>0.8$, there are some discrepancies and both
the frequency shift and damping rate simulations deviate from the
observed behaviour.
\begin{figure}
\begin{center}\mbox{ \epsfxsize 4in\epsfbox{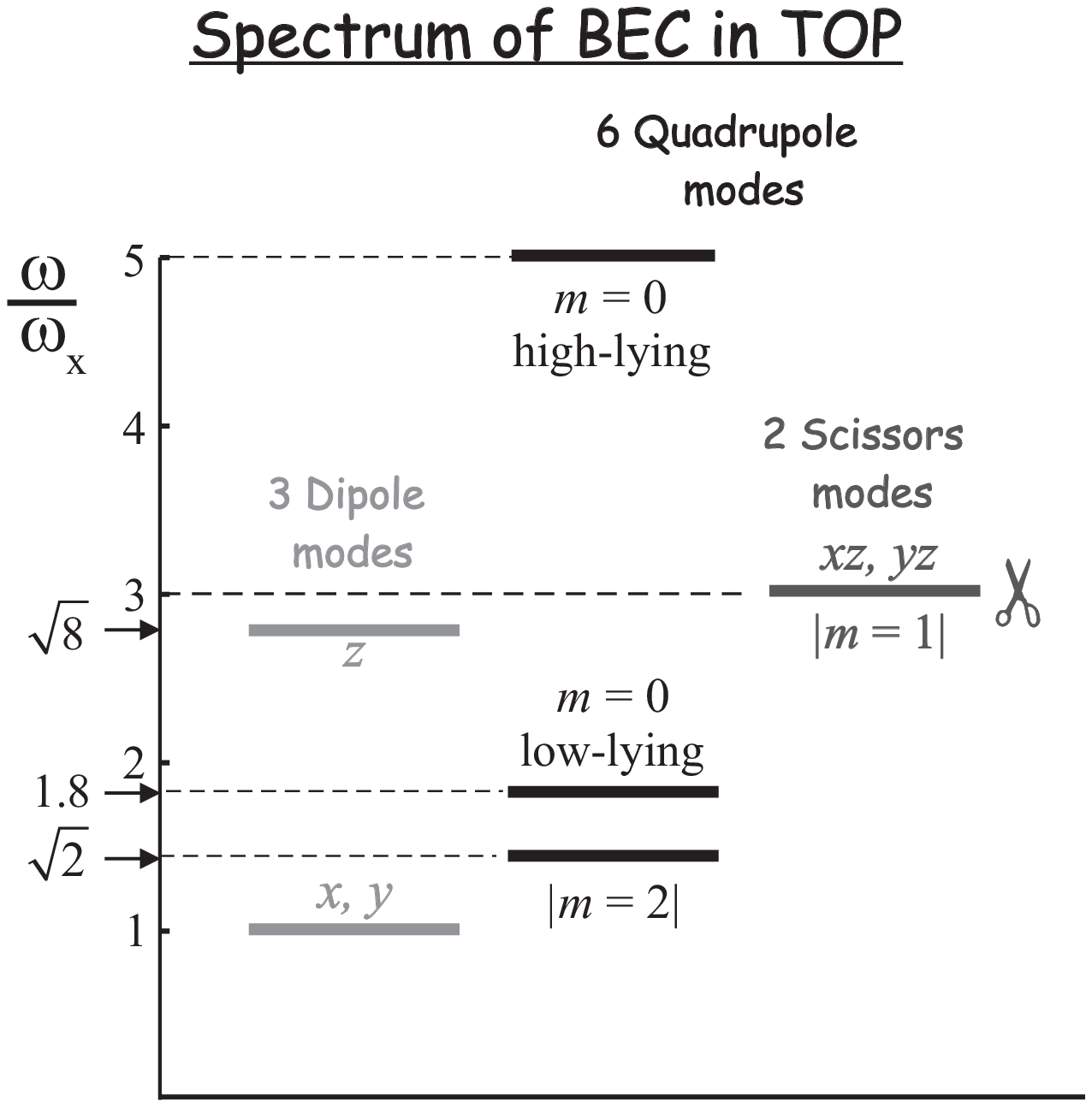}}\end{center}
\caption{Low energy spectrum of a trapped BEC in a TOP trap. In a
TOP trap geometry $\omega_x=\omega_y$, the $z$ component of the
angular momentum is conserved and $m$ is a good quantum number.
There are three dipole modes: one at $\omega_z$ and two degenerate
modes at $\omega_x$. The six quadrupole modes are: $m=0$
high-lying, $m=0$ low-lying, the doubly degenerate $|m=2|$ and two
degenerate scissors modes. In an anisotropic trap we have three
dipole modes, three scissors modes (odd parity) and three
compressional modes (even parity).}\label{spectrum-1}
\end{figure}

The scissors mode is a previously unobserved collective mode of a
BEC and Fig.~\ref{spectrum-1} shows how it fits in with the other
energy modes. The dipole modes (center of mass motion) of a BEC
occur at the trap frequencies by Kohn's theorem, and we used this
property to calibrate our trap frequencies. The frequencies of
other collective modes do not directly correspond to harmonic
oscillator states because of the atomic interactions. In an
axially symmetric trap the angular momentum about the symmetry
axis is conserved and we use its eigenvalue $m$ to label the low
energy modes. We have a high and low-lying $m=0$ modes and the
$m=2$ mode~\cite{Jin1996a}. Moreover there are two degenerate
($\omega_x=\omega_y$) scissors modes with frequencies
$\omega_{xz}=\omega_{yz}=\omega_{sc}$.

\section{MOMENT OF INERTIA}
The moment of inertia is a fundamental quantity of a physical
system that characterizes its rotational properties. For rotation
about the z axis a classical rigid body has a moment of inertia
given by $\Theta_{rig} = m N \langle x^2 + y^2 \rangle$ where $N$
is the number of particles in the system, $m$ their mass and
$\langle \rangle$ denotes an ensemble average. In superfluids the
response to a weak rotational field is always less that of a
classical rigid body. This {\it quenching} of the moment of
inertia have been studied in different many-body systems. In early
experiments on liquid helium by Andronikashvili, measurements on
the moment of inertia gave a quantitative measure of the
superfluid density~\cite{Huang}. In nuclear physics the strong
reduction of the moment of inertia with respect to the rigid body
value gave indication of the superfluidity of nuclear matter soon
after the discovery of nuclear rotational states~\cite{Bohr75}.

In the framework of linear response theory~\cite{Nozieres} the
moment of inertia of a gas $\Theta$ is a dynamical property of the
system. Relative to a certain rotation axis, it is defined as the
linear response to a rotational probe field $-\Omega \hat{J}_z$
\begin{equation}
\Theta=\lim_{\Omega \rightarrow 0} \frac{\langle \hat{J}_z
\rangle}{\Omega}.
\end{equation}
For a Bose condensed gas this definition holds for systems
rotating with angular velocities less than the critical value
$\Omega_c$ which gives the occurrence of quantized vortices. In
our experiments we study the response to an angular impulse and
the {\it rotational probe} is the scissors excitation. The link
between the scissors mode and the moment of inertia has been
recently established by Zambelli and
Stringari~\cite{Zambelli2001}. In their analysis they calculate
explicitly the moment of inertia of the boson gas $\Theta$ using
the quadrupole response function. This derivation leads to the
evaluation of $\Theta$ for the condensate at $T=0$ and the thermal
cloud at $T>T_0$ directly from scissors mode frequency
measurements.

Following their approach we can write the Hamiltonian of an
interacting boson gas as the sum of the kinetic energy, a term
that represents the two-body interaction between the particles and
the confining harmonic potential $\hat{V}_{ext}(\vec{r})$ (the
operator for the potential of Eq.~\ref{potential}). The crucial
point of their analysis is the commutation relation that follows
from the rotational symmetry of the internal part of the
Hamiltonian (kinetic and self-interaction terms), so that:
\begin{equation}
\left[ \hat{H}, \hat{J}_z \right] = \left[ \hat{V}_{ext},
\hat{J}_z \right] = i m \hbar (\omega_x^2 - \omega_y^2) \hat{Q}
\label{commute}
\end{equation}
\noindent where $\hat{Q}= x y$ is the quadrupole
operator~\cite{note}. This equation links the dynamics of the
angular momentum with the quadrupole operator that arises because
of the deformation of the trap. From this important relation and
from the definition of moment of inertia in the context of linear
response theory it is possible to write $\Theta$ as a function of
the imaginary part of the quadrupole dynamic response
function~\cite{Nozieres}. Thus the occurrence of superfluidity
that is responsible for the quenching of the moment of inertia is
fundamentally linked to the dynamics of quadrupole excitations.

In a scissors mode experiment we have a sudden rotation of the
trap through a small angle $\theta_0$. The resulting external
time-dependent Hamiltonian comes from the angular displacement of
the potential with respect to the atomic cloud and for $t>0$ can
be written as $\hat{H}_{ext}(t) = m\theta_0 (\omega_x^2
-\omega_y^2) \hat{Q}$. Using linear response theory it is then
possible to find the evolution of the average $Q(t) \equiv \langle
\hat{Q} \rangle$ and a relation between the reduced moment of
inertia and the quadrupole Fourier signal
$Q(\omega)$~\cite{Zambelli2001}:
\begin{equation} \Theta^{\prime}=(\omega_x^2-\omega_y^2)^2
\frac{\int d\omega\, Q(\omega)/\omega^2}{\int d\omega \,
Q(\omega)\omega^2}.\label{fourier}
\end{equation}
This important general relation allows us to extract information
about the moment of inertia from the Fourier transform of the
quadrupole moment. This signal in our experiment is intrinsically
related to the scissors mode angle oscillation $Q(t)=N\langle y^2-
x^2 \rangle \theta(t) \label{q2}$.

The relations discussed so far are very general and do not rely on
any model for the cloud density distribution. We can apply these
equations in different regimes and specifically to our scissors
mode data. At $T=0$ the condensate has a single frequency scissors
oscillation both for the ideal and the interacting gas. From this
prediction we have that the quadrupole signal has a delta-like
Fourier spectrum with amplitude $Q_{0,c}=\theta_0 N\langle y^2-
x^2 \rangle_c $ and resonant frequency $\omega_{sc}(N,T)$.
Application of Eq.~\ref{fourier} yields:
\begin{equation}
\Theta_c^{\prime}=\frac{(\omega_x^2-\omega_y^2)^2}{\omega_{sc}^4(N,T)}.\label{inertCond}
\end{equation}

\begin{figure}
\begin{center}\mbox{ \epsfxsize 5.8in\epsfbox{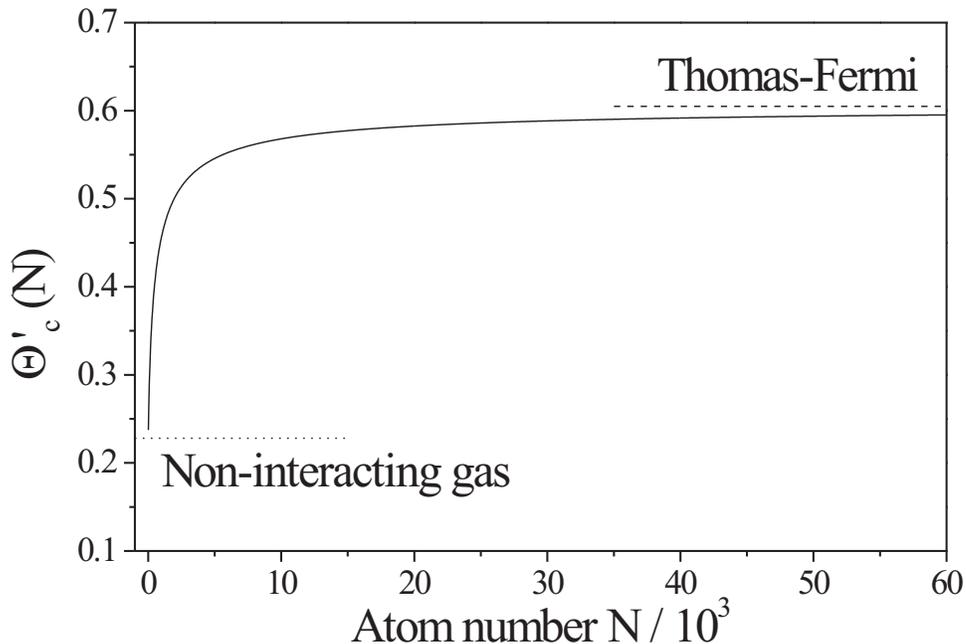}}\end{center}
\caption{Moment of inertia and atom number. For very small number
of atoms the moment of inertia is consistent with the $T=0$
non-interacting gas result (dotted line) i.e.
$\Theta_{ho}^{\prime}=\epsilon_{ho}^2\approx 0.23$. In the large
atom number limit the moment of inertia approaches the
Thomas-Fermi irrotational value (dashed line) i.e.
$\Theta_c^{\prime}=\epsilon^2\approx 0.6$.} \label{inertnum}
\end{figure}

In the specific case of the TOP trap geometry we have that
$\omega_x=\omega_y$ and so $\Theta^{\prime}_c$ with respect to z
is zero due to the rotational symmetry. The axial frequency is
$\omega_z=\sqrt{8} \omega_x$ and we can consider a scissors
oscillation about the y axis. In the Thomas-Fermi limit, using the
expression for the scissors mode frequency defined in
Sec.~\ref{scistheory}, we obtain the quenched irrotational value
of the moment of inertia $\Theta_c^{\prime}=\epsilon^2\simeq 0.6$.
We can also use Eq.~\ref{inertCond} for a non-interacting
condensate. In this case the scissors frequency is simply the
harmonic oscillator prediction $\omega_{sc,ho}= \omega_z
+\omega_x$ and we eventually obtain the ideal gas value of the
moment of inertia~\cite{Stringari1996b}
$\Theta_{ho}^{\prime}=\epsilon_{ho}^2\simeq 0.23$, where
$\epsilon_{ho} = (\omega_x-\omega_y)/(\omega_x+\omega_y)$ is the
deformation of the harmonic oscillator ground state. Indeed the
interaction among the particles changes dramatically the
rotational properties of the BEC embodied in the dynamics of the
scissors mode. Therefore to evaluate this effect we can
investigate the frequency change of the scissors mode with atom
number~\cite{Pires2000} and make use of Eq.~\ref{inertCond}. The
result of this procedure is shown in fig.~\ref{inertnum}. These
calculations were made for a TOP trap with radial frequency
$\omega_x/2\pi=126$~Hz. For very low atom number, i.e. low
interactions strength, we recover the ideal gas result
$\Theta_{ho}^{\prime}\simeq 0.23$. While in the limit of large $N$
the moment of inertia approach the Thomas-Fermi result
$\Theta_c^{\prime}\simeq 0.6$.

For a collisionless thermal cloud the response to a scissors
excitation exhibits two frequencies at $\omega_+$ and $\omega_-$
with the same amplitude $\theta_0/2$. The quadrupole signal
amplitudes are related to the angle amplitude through
$Q_{0,th}=N\langle y^2- x^2 \rangle_{th} \theta_0/2$. Application
of Eq.~\ref{fourier} yields the moment of inertia for the gas
above $T_0$:
\begin{equation}
\Theta_{th}^{\prime}=\frac{(\omega_x^2-\omega_y^2)^2}{\omega_+^2\
\omega_-^2}.\label{inertTherm}
\end{equation}
Using the scissors mode collisionless frequencies we obtain the
rigid body value for the moment of inertia $\Theta^{\prime}=1$ as
expected for a classical gas. For the thermal cloud the occurrence
of the low frequency mode at $\omega_-$ is the key feature that
yields the rigid body response of the gas to the scissors
excitation. For both $T=0$ and $T>T_0$ the reduced moment of
inertia only depends on the scissors mode frequencies and the
geometry of the trap - it is independent of the amplitude of the
oscillations. This is not the case in the intermediate regime i.e.
for condensed clouds at finite temperature.

In the general case when the Bose condensed cloud exhibits a
scissors motion at finite temperature the analysis of the {\it
total} moment of inertia through the scissors oscillation becomes
more complicated. Indeed the normal and superfluid component
respond in a different way to the sudden rotation of the trap.
However we can extend the formalism developed in
\cite{Zambelli2001,Stringari1996b} to extract important
information on the total moment of inertia in a semi-classical
approximation. This approximation holds when the temperature $T$
of the system is much larger than the harmonic oscillator energy
$\hbar \omega_{ho}$. This is the case for current experiments on
dilute gases and in particular for our experiment.

First of all for partially condensed clouds we need to introduce
two distinct angles for the condensate $\theta_c(t)$ and the
thermal cloud $\theta_{th}(t)$. Thus after the sudden rotation of
the trap the time dependent density of the gas will have a time
dependent contribution from both the condensed and the thermal
components. We can use the semi-classical approach to evaluate the
total quadrupole moment of the gas. In fact, in the linear regime
$Q(t)$ is the sum of two terms, one related to the BEC and one
related to the thermal cloud oscillation $Q(t)=Q_c(t)+Q_{th}(t)$
with
\begin{eqnarray}
Q_c(t)&=&N_0(T) \langle y^2 - x^2 \rangle_c \theta_c (t) \\
Q_{th}(t)&=& \left[ N-N_0(T) \right] \langle y^2-x^2 \rangle_{th}
\theta_{th}(t).
\end{eqnarray}
Where $N$ is the total number of atoms, $N_0(T)$ is the
temperature dependent number of condensed atoms, and the average
$\langle \rangle_{c}$ and $\langle \rangle_{th}$ are evaluated
over the condensate and thermal distributions respectively. The
total Fourier signal will be the sum of two distinct terms as
well. In particular the condensate responds at one frequency for
all temperatures. Instead the thermal cloud signal is a two
frequency oscillation at all temperatures (see fig.~\ref{freq}).
Neglecting the damping we can we can write the Fourier quadrupole
signal as a superposition of $\delta$-like functions with
temperature dependent amplitudes:
\begin{eqnarray}
 &Q_{0, th}(T)=\frac{\theta_0 k_B T}{2m} \frac{G(3)}{3G(2)}
\left[ N-N_0(T) \right]
\frac{\omega_x^2-\omega_y^2}{\omega_x^2\omega_y^2}&
\label{qzeroth}
\\
 & Q_{0,c}(T)=
\frac{2\theta_0 \mu(T) N_0(T)}{7 m}\, \frac{\omega_x^2-
\omega_y^2}{\omega_x^2\omega_y^2}.& \label{qzeroc}
\end{eqnarray}
where $G(3)\approx6.5$, $G(2)\approx 2.4$ are numerical
coefficients~\cite{Stringari1996b}.
Substituting the total Fourier signal $Q(\omega)$ in
Eq.~\ref{fourier} under the conditions outlined above, we get:
\begin{equation}
\Theta^{\prime}(T) = \frac{\Theta_c^{\prime}(T) +
\Theta_{th}^{\prime} {\mathcal F}(N,T)}{1+ {\mathcal
F}(N,T)}.\label{totalinert}
\end{equation}
Here $\Theta_c^{\prime}(T)$ and $\Theta_{th}^{\prime}$ are the
same as Eqs.~\ref{inertCond} and \ref{inertTherm} respectively but
with the temperature dependent frequencies of the scissors mode.
Instead the temperature dependent function ${\mathcal F}(N,T)$ is
related to the signal amplitudes and takes the expression:
\begin{equation}
{\mathcal F}(N,T)=\frac{Q_{0,th}}{Q_{0,c}}\, \frac{\omega_+^2
+\omega_-^2}{\omega_{sc}^2}.
\end{equation}
In the Thomas-Fermi limit we eventually obtain:
\begin{equation}
{\mathcal F}(N,T)=\frac{7 k_B T}{4 \mu (T)}\frac{G(3)}{3 G(2)}
\frac{N-N_0(T)}{N_0(T)} \frac{\omega_+^2
+\omega_-^2}{\omega_{sc}^2(T)}.\label{new-stuff-important}
\end{equation}
It is interesting to consider the two limiting cases for
$T\rightarrow 0$ and $T\rightarrow T_0$. In the first situation we
have that $\lim_{T\rightarrow 0} {\mathcal F}(N,T)=0$ and
$\Theta^{\prime}(0)=\Theta_c^{\prime}(0)=\epsilon^2$. While in the
limit $T\rightarrow T_0$ the function ${\mathcal F}(N,T)$ diverges
and we recover the rigid body value of the moment of inertia
$\Theta^{\prime}(T_0)= \Theta_{th}^{\prime}=1$. This result can be
compared with the one obtained from the densities without
considering any dynamics of the clouds~\cite{Stringari1996b}:
\begin{equation}
\Theta^{\prime} (T)=\frac{\epsilon^2 +
f(N,T)}{1+f(N,T)}\label{new-stuff}
\end{equation}
where the function $f(N,T)$ is the ratio between the rigid body
moments of inertia for thermal cloud and condensate and it depends
only on condensate fraction, chemical potential and temperature of
the Bose gas:
\begin{equation}
f(N,T)=\frac{\Theta_{th,rig}}{\Theta_{c,rig}}=\frac{7 k_B T}{2 \mu
(T)}\frac{G(3)}{3 G(2)}\frac{N-N_0(T)}{N_0(T)}.
\end{equation}
Thus $f(N,T)$ resembles ${\mathcal F}(N,T)$ in
Eq.~\ref{new-stuff-important} apart from the ratio of the squares
of the scissors mode frequencies at $T=0$ to that at $T$. This
ratio differs significantly from unity only at temperatures $T
\geq 0.6 T_0$ where the thermal cloud dominates the contribution
to the total moment of inertia. So we expect the static results to
give a good description of the moment of inertia of the gas at all
temperatures.

We can now use the analytical results obtained so far to deduce
the total moment of inertia of the partially condensed system from
our experiments on the scissors mode. First we can use
Eqs.~\ref{inertTherm} and \ref{inertCond} to get the moment of
inertia for the thermal and condensate {\it separately} from the
finite temperature frequency shifts of the scissors oscillations.
\begin{figure}
\begin{center}\mbox{ \epsfxsize 5in\epsfbox{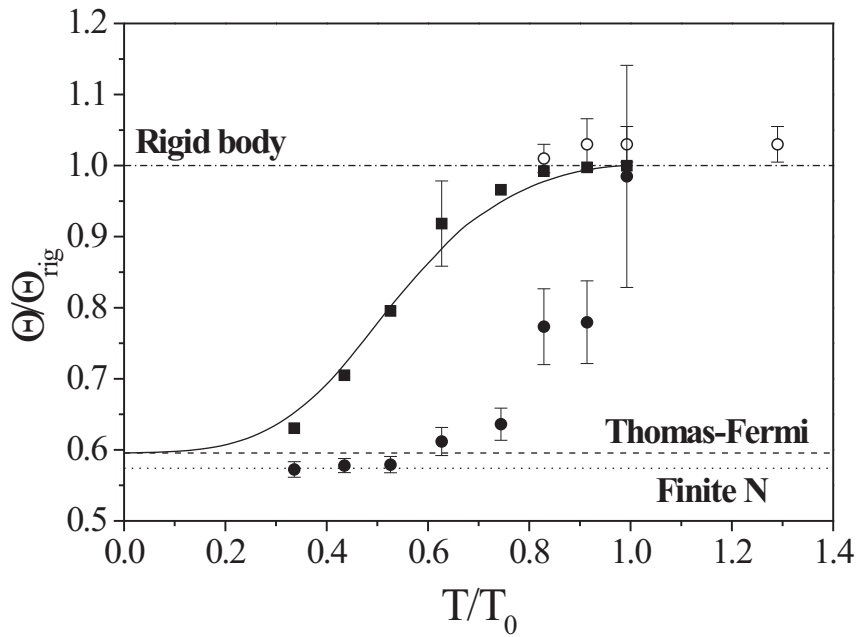}}\end{center}
\caption{Reduced moment of inertia at finite temperature.
Temperature dependence of the moment of inertia of the thermal
(open circles) and condensed component (solid circles). The errors
on these data points reflect the relative errors on the scissors
mode frequencies. For low temperatures the moment of inertia of
the condensate is in close agreement with the hydrodynamic
prediction for our geometry (dashed line) and it is in excellent
agreement with the finite number correction (dotted line). The
solid line is the prediction of Eq.~\ref{new-stuff} that is based
only on an estimate of the cloud widths of the interacting gas and
neglects the dynamical interaction between the BEC and the thermal
cloud. Finally the solid squares and line represent the moment of
inertia for the total quantum gas extracted by using
Eqs.~\ref{totalinert} and \ref{new-stuff-important} and the
scissors mode data. The error on these data points is of the order
of $7\%$.} \label{exp-inertia}
\end{figure}
In Fig.~\ref{exp-inertia} we show the behaviour of
$\Theta^{\prime}_c(T)$ (solid circles) and
$\Theta^{\prime}_{th}(T)$ (open circles) with temperature. Note
that the moment of inertia of the thermal component is consistent
with the rigid body value for all measured temperatures in the
experiment $\Theta^{\prime}_{therm}\approx 1$. We can see that all
the relevant quantities involved in Eqs.~\ref{totalinert} and
\ref{new-stuff-important} for the evaluation of the moment of
inertia can be extracted from TOF images~\cite{thesis}. The total
atom number $N$ and condensate number $N_0(T)$ are obtained from a
2D double distribution: parabolic for the BEC and Gaussian for the
thermal cloud. The temperature of the cloud is deduced by fitting
the wings of the thermal cloud with a 2D Gaussian distribution.
The chemical potential has been deduced using the Castin-Dum
equations for the expanding BEC. Using all the data extracted from
the images we can `construct' the moment of inertia of the total
boson gas from the quadrupole Fourier signal
(Eq.~\ref{totalinert}) for each measured temperature. The result
is shown in Fig.~\ref{exp-inertia} (solid squares) and we can
compare it with the thermal and condensate contribution. The
hydrodynamic prediction (dashed line) and the finite number
correction (dotted line) are also shown for comparison. A theory
for the scissors mode frequency has been only recently developed
and also used to extract information on the moment of inertia at
finite temperature~\cite{JacksonZaremba2001}. Our results are
consistent with a monotonic increase of the moment of inertia with
increasing temperature and close to $T_0$ the condensate scissors
mode frequency correspond to an effective moment of inertia close
to the rigid body value. We can also compare our approximation for
the total moment of inertia with the estimate obtained from
Eq.~\ref{new-stuff} using the condensate fraction for the
interacting gas~\cite{anna}. This is shown as the solid line in
Fig.~\ref{exp-inertia}. Although we are neglecting the BEC
scissors mode temperature frequency shift i.e. the dynamical
interaction between the thermal cloud and BEC component, the
agreement between the extracted data points and the static theory
is good.

\begin{figure}
\begin{center}\mbox{ \epsfxsize 5.8in\epsfbox{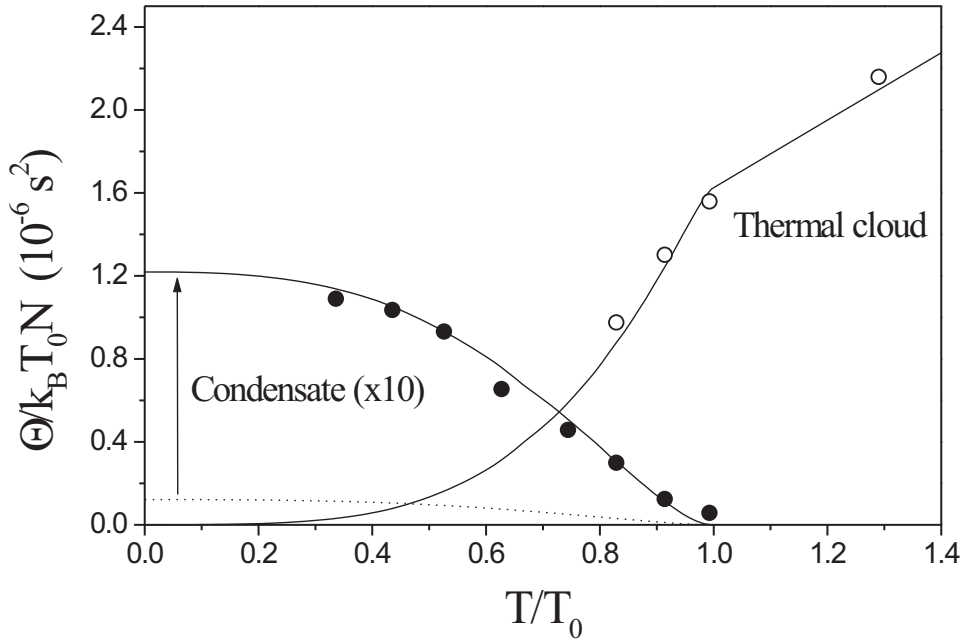}}\end{center}
\caption{Temperature dependence of the absolute moment of inertia
of the thermal cloud (open circles) and condensate (solid
circles). The solid lines are the interacting theory predictions.
The condensate data and theory have been multiplied by a factor of
$10$ to make them more visible. The dotted line is the original
theoretical prediction for the BEC moment of inertia. The
condensate and thermal component have equal moment of inertia at
$T\approx 0.47\, T_0$.} \label{not-reduced}
\end{figure}
So far we have investigated the behaviour of the reduced moment of
inertia of the gas. It is also interesting to present the data in
terms of the absolute (not reduced) values of the moments of
inertia for the two components of the system i.e.
$\Theta_c=\Theta_c^{\prime} \Theta_{c,rig}$ and
$\Theta_{th}=\Theta_{th}^{\prime} \Theta_{th,rig}$. In the large
number limit we obtain:
\begin{eqnarray}
\frac{\Theta_{c}}{Nk_B T_0}&=& \Theta_{c}^{\prime} \,
\frac{N_0}{N}\frac{2\mu}{7k_BT_0}\frac{\omega_x^2
+\omega_y^2}{\omega_{x}^2\omega_y^2}\\ \frac{\Theta_{th}}{Nk_B
T_0}&=&\Theta_{th}^{\prime} \frac{T}{T_0} \,
\frac{N-N_0}{N}\frac{G(3)}{3 G(2)}\frac{\omega_x^2
+\omega_y^2}{\omega_{x}^2\omega_y^2}
\end{eqnarray}
In particular as a consequence of the harmonic confinement, the
temperature dependence of the moment of inertia for the thermal
component is linear above $T_0$ and behaves as $(T/T_0)^{4}$ below
$T_0$. Figure~\ref{not-reduced} shows the experimental data points
and predictions based on the interacting theory for both
condensate and thermal cloud absolute moment of inertia. The
condensate and thermal cloud have equal moments of inertia
$\Theta_c=\Theta_{th}$ at $T\approx 0.47\, T_0$. At this
temperature only $15\%$ of the atoms are in the thermal cloud, but
the cloud has a classical moment of inertia and a broader spatial
distribution than the BEC.

\section{CONCLUSIONS}
In conclusion we have shown how the scissors mode is an excellent
rotational probe to study the superfluid properties of a quantum
gas. Specifically we have outlined its link with the moment of
inertia of the gas. This allowed us to explicitly measure the
quenching of the moment of inertia from our scissors mode data.

Another system where the scissors mode could be used as a powerful
tool to demonstrate superfluid flow is a Fermi degenerate gas. A
recent theoretical analysis~\cite{MinguzziFermi} of that system
shows that the transition to the superfluid state, below the BCS
critical temperature $T_{BCS}$, is signaled by the disappearance
of the low frequency in the scissors oscillation that occur in the
normal degenerate Fermi gas above $T_{BCS}$. The same analysis
used here for extracting the moment of inertia of the gas can be
performed for a trapped Fermi gas to directly measure the
quenching of its moment of inertia below $T_{BCS}$. In this case
we have a situation similar to heavy nuclei and some theoretical
predictions have been already presented in \cite{Farine2000}.

\section{ACKNOLEDGEMENTS}
We would like to thank S. Stringari, F. Zambelli and N. Lo Iudice
for fruitful discussions. We are also grateful to D. Cassettari
and S. Cornish for comments and proof reading of the manuscript.
This work was supported by the EPSRC and the TMR `Cold Quantum
Gases' network program (No. HPRN-CT-2000-00125). O.M. Marag\`{o}
acknowledges the support of a Marie Curie Fellowship, TMR program
(No. ERB FMBI-CT98-3077) and Linacre College.


\begin{thebibliography}{9}
\bibitem{note}
The angular momentum evolution for a quantum system is in general
given by the Heisenberg equation of motion $d \vec{J}/dt = -i
\left[ \vec{J}, H \right]/\hbar$. During a scissors mode
oscillation the angular momentum is intrinsically linked to the
quadrupole moment. Thus if we excite the xz scissors mode, then
using relation \ref{commute}, we can write the evolution of the
relevant averaged component $\langle J_y \rangle$ of the angular
momentum $d\langle J_y \rangle/dt = -m (\omega_z^2 -
\omega_x^2)\langle Q \rangle=-m N (\omega_z^2 - \omega_x^2)\langle
x^2- z^2 \rangle \theta(t)$. From the scissors mode angle
evolution we can easily integrate the previous equation and using
the Thomas-Fermi mean square widths we finally get the explicit
evolution of the angular momentum: $$ \langle J_y \rangle = -
\frac{2\theta_0\mu N}{7\omega_{sc}} \frac{\left( \omega_z^2
-\omega_x^2\right)^2}{\omega_x^2\omega_z^2} \sin (\omega_{sc}
t).$$ During a scissors mode oscillation the angular momentum
oscillates at the scissors mode frequency and out of phase with
respect to the scissors mode angle.
\end{thebibliography}
\end{document}